# Preparation and Characterization of α-Fe$_2$O$_3$ polyhedral nanocrystals via annealing technique

Rajesh Kumar[1*], S. Gautam[2$], In-Chul Hwang[3], K. H. Chae[2] and Nagesh Thakur[1]

[1]Department of Physics, Himachal Pradesh University, Shimla 171-005, India

[2]Materials Science and Technology Research Division, Korea Institute of Science and Technology (KIST), Seoul 136-791, Korea

[3]Department of Chemistry, Pohang University of Science and Technology (POSTECH), Pohang 790-784, Korea

## Abstract

Polyhedral nanocrystals of α-Fe$_2$O$_3$ are successfully synthesized by annealing FeCl$_3$ on silicon substrate at 1000 $^o$C in the presence of H$_2$ gas diluted with argon (Ar). Uniformly shaped polyhedral nanoparticles (diameter ~50-100 nm) are observed at 1000 $^o$C and gases flow rate such as; Ar = 200 ml/min and H$_2$ = 150 ml/min. Non-uniform shaped nanoparticles (diameter ~ 20-70 nm) are also observed at an annealing temperature of 950 $^o$C with lower gases flow rate (Ar = 100 ml/min and H$_2$ = 75 ml/min). Nanoparticles are characterized in detail by field-emission electron microscopy (FE-SEM), energy dispersive x-ray (EDX) and high resolution transmission electron microscopy (HRTEM) techniques. HRTEM study shows well resolved (110) fringes corresponding to α-Fe$_2$O$_3$, and selected area diffraction pattern (SADP) confirms the crystalline nature of α-Fe$_2$O$_3$ polyhedral nanoparticles. It is observed that polyhedral formation of α-Fe$_2$O$_3$ nano crystals depends upon annealing temperature and the surface morphology highly rely on the gas flow rate inside the reaction chamber.

$Corresponding authors:

E-mail: sgautam71@kist.re.kr (S. Gautam); rajeshkumarf11@gmail.com(Rajesh Kumar)

Ph: +82-54-279-1503; Fax: +82-54-279-1599

---

*Also at Department of Chemistry, Pohang University of Science and Technology, Pohang 790-784, Korea
$Also at Pohang Accelerator Laboratory, Pohang University of Science and Technology, Pohang 790-784, Korea.

## 1. Introduction

Recently tremendous interest has been generated in the study of metal/metal oxides nanoparticles because they lead to a new generation of optics, electronics, sensors and photocatalysis [1,2]. Among them magnetic nanoparticles have been a subject of increasing interest due to their interesting physical properties and different technological applications [3,4]. During past few years, magnetic nanomaterials have been the focus of the research, and the magnetic metals, alloys and metal oxides have lead to great advances in the formation of different nano structures [5-10]. Iron oxide nanoparticles are of considerable interest because of the valuable applications such as magnetic storage, medicine, and as catalyst [11,12]. Over the years researchers have employed in different routes to facilitate large-scale synthesis of iron oxide nanoparticles using different synthesis techniques such as sol–gel processes[13], chemical precipitation[14,15] forced hydrolysis[16,17] etc.[18-21]. However, the temperature effect and gas flow during synthesis, on the structure and surface morphology of the nanoparticles have received less attention till date. Here we report the synthesis of uniform polyhedral nanocrystals of α-$Fe_2O_3$, in which suitable growth conditions have been facilitated to form polyhedral form of α-$Fe_2O_3$ nanocrystals.

## 2. Experimental

Synthesis of polyhedral α-$Fe_2O_3$ nanoparticles were carried out with a simple annealing method. A Si (100) substrate was cleaned by hydrofluoric acid solution (5%) in order to remove the pre-existing native oxide layer. The substrate was rinsed with deionized water and dried with a stream of nitrogen gas. A drop of 2M (mole) solution of $FeCl_3$ in ethanol and water (1:1 by volume) was placed on the cleaned Si substrate, and was then transferred to the quartz tube (reaction chamber) inside a furnace. The argon (Ar) gas was allowed to flow at a rate of 100 ml/min into the reaction chamber immediately after loading the sample and it was continued throughout the annealing process. For annealing, the temperature of the sample was raised to 950 $^o$C (at a heating rate of 15.83 /min) and maintained at 950 $^o$C for 15 minutes. During the increase of temperature hydrogen ($H_2$) gas was introduced into the quartz tube at a rate of 75 ml/min in order to cause the reduction of $FeCl_3$ on the Si substrate. After annealing the sample for 15 minutes at 950 $^o$C, the furnace was cooled to the room temperature at a normal cooling rate (~1 $^o$C/min). In another experiment, for the same annealing time (15 min) the temperature was increased to 1000 $^o$C and the gas flow was increased to as; Ar = 200 ml/min, and $H_2$ = 150 ml/min. The obtained samples were characterized by FE-SEM (Philips, XL30S, working at 5kV), EDX installed in FE-SEM, and HRTEM (JEOL, JEM-2100F, performing at 200 kV) installed with EDX.

## 3. Results and discussion

Figure 1(a) shows the top view of FE-SEM image of nanoparticles synthesized on Si substrate at annealing temperature 950 $^o$C, and flow rate such as Ar = 100 ml/min and $H_2$ = 75 ml/min. A large number of nanoparticles are clearly observed on the Si substrate with diameters in range of ~ 20 -70 nm, as shown in the figure 1(a). In the high magnification FE-SEM image shown in figure 1(b) the particles can be observed to be composed of many rough surfaces and having non-uniform shapes. On the other hand, the nanoparticles formed at 1000 $^o$C for higher values of gas flow (Ar = 200 ml/min and $H_2$ = 150 ml/min) as shown in the figure 2(a) are polyhedral nanoparticles and having average diameters range of 50 nm - 100 nm. The nanoparticles obtained at 1000 $^o$C are slightly larger in size as compared to those obtained at 950 $^o$C. High magnification FE-SEM image for a nanoparticle obtained at 1000 $^o$C is shown in the figure 2(b), which is a uniform polyhedral shape. Thus, the nanoparticles obtained at two different temperatures and gas flow rates are clearly distinguishable in shape and surface morphology as shown in figures 1 and 2 respectively. . It is suggested that the annealing temperature and gas flow inside the reaction chamber significantly affect the shape and surface morphology of the nanoparticles. In high magnification FE-SEM image as shown in the figure 2(b), the small nanoparticles can be seen at the surface of bigger polyhedral nanoparticle. The formation of polyhedral nanoparticle can be treated as a self arranged form of the small nanoparticles at annealing temperature 1000 $^{o}$C, while at lower temperature (950 $^{o}$C) non-uniform shaped nanoparticles have been observed. It has been inferred that 950 $^{o}$C may not be sufficient temperature for the self arrangement of small nanoparticles into polyhedral form. In other words, with the increase of annealing temperature small nanoparticles sink together and a big nanoparticle gradually emerges in polyhedral form as a result of self arrangement of small nanoparticles. This sinking behavior of small nanoparticles into bigger particle is clearly observed by high resolution FE-SEM image as shown in the figure 2(b), where many small nanoparticles are visible in spreaded form at the surfaces of big nanoparticle. Further in case of low flow rate of gas inside reaction chamber rough surfaces of nanoparticles are observed in high magnification FE-SEM image (figure 1b), while with increasing gas flow rate at 1000 $^{o}$C the surface of nanoparticles has been observed comparatively smooth, as shown in figure 2(b). Based upon these FE-SEM observations, the change in structure of nanoparticles (non-uniform to uniform shape) is connected to the annealing temperature and the change in surface morphology is connected to the flow rate of Ar and $H_2$ gases inside the reaction chamber. The high annealing temperature facilitates the self assembly of small nanoparticles into polyhedral nanoparticles and the high flow rate of gases causes smoothness of their surfaces.

In order to know the elemental composition of the polyhedral nanostructure, EDX analysis was carried out as shown in the figure 2(c). The EDX analysis (installed in FE-SEM) of the polyhedral nanoparticle represents the atomic percentage (%) of the elements such as Fe, O and Si in the ratio of ~ 2:5:1 (table in the figure 2). The Si signal in EDX spectra was expected from the silicon substrate itself, as the EDX analysis of polyhedral nanoparticles was carried out on Si substrate, therefore the contribution of silicon (Si) to the nanocrystal could not be justified by EDX analysis in the FE-SEM study. The absence of Si in the nanocrystals was further confirmed by EDX analysis installed in TEM, where only Fe and O was found in a ratio of 2:3 forming $Fe_2O_3$ nanoparticles. From the TEM image shown in the figure 3(a) of polyhedral nanocrystal formed at 1000 °C (Ar = 200 ml/min and $H_2$ = 150 ml/min), the average length of edge of the polyhedral nanoparticle is measured to be 50 nm. Also in high resolution TEM (HRTEM) image as shown in the figure 3(b), a fringe type pattern is observed with a lattice spacing 0.252 nm. The observed interplaner spacing for the nanostructure in HRTEM study was also confirmed by the atomic profile measurements as shown in the figure 3(c). The observed value of interplaner spacing corresponds to 110 planes (with lattice spacing 0.251 nm) of the α-$Fe_2O_3$ structure. Moreover, the selected area electron diffraction (SADP) pattern in the figure 3(d) taken from the polyhedral nanoparticles show that the particle is a single crystal. Therefore, from HRTEM analyses the structure formed at 1000 °C is confirmed as the α-$Fe_2O_3$.

## 4. Conclusions

In summary, we have successfully synthesized α-$Fe_2O_3$ polyhedral nanocrystals of diameter ~50-100 nm on a Si substrate via a simple annealing technique. The change of non-uniform shaped nanoparticles to uniformly shaped polyhedral nanoparticles with the annealing temperature and the gas flow rate inside the reaction chamber are observed. The formation of polyhedral form of α-$Fe_2O_3$ nanocrystal has been found to be dependent on the temperature, while the surface morphology highly rely on the gas flow rate inside the reaction chamber. A polyhedral nanocrystal of α-$Fe_2O_3$ is observed to be composed of many small nanoparticles.

**Acknowledgements**

This work is supported by CSM Lab department of Chemistry, POSTECH, Korea. SG and KHChae are thankful to KIST support (Grant No. 2V01320).

**Figure Captions**

**Figure 1.** FE-SEM image of non-uniform shaped nanoparticles synthesized at 950 $^{o}$C, on the Si (100) surface. (a) Top view of the nanoparticles. (b) A high magnification image at a viewing angle of 45$^{o}$.

**Figure 2.** FE-SEM image of the $Fe_2O_3$ polyhedral nanocrystals obtained at 1000 $^{o}$C annealing temperature. (a) The nanocrystals are polyhedral with different smooth surfaces. (b) High resolution FE-SEM image clearly indicating different faces of polyhedral nanocrystal. (c) Represents the EDX spectrum with atomic % in the table.

**Figure 3.** (a) TEM image, a $Fe_2O_3$ polyhedral nanocrystal. (b) High resolution TEM image showing parallel 110 planes. (c) Atomic profile for the HRTEM image. (d) Selected area diffraction pattern (SADP) of the nanocrystal.

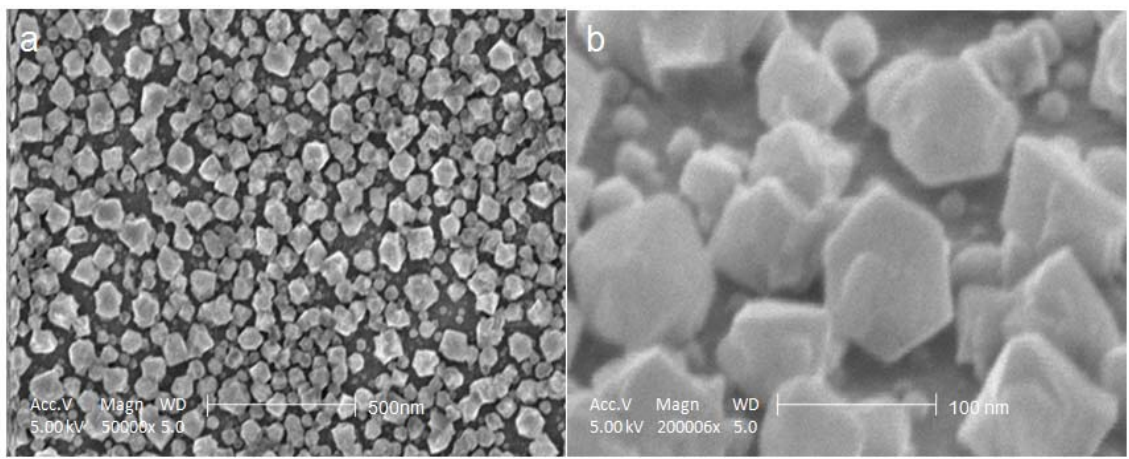

**Figure 1: FE-SEM image of non-uniform shaped nanoparticles synthesized at 950 $^{o}$C, on the Si (100) surface. (a) Top view of the nanoparticles. (b) A high magnification image at a viewing angle of 45$^{o}$.**

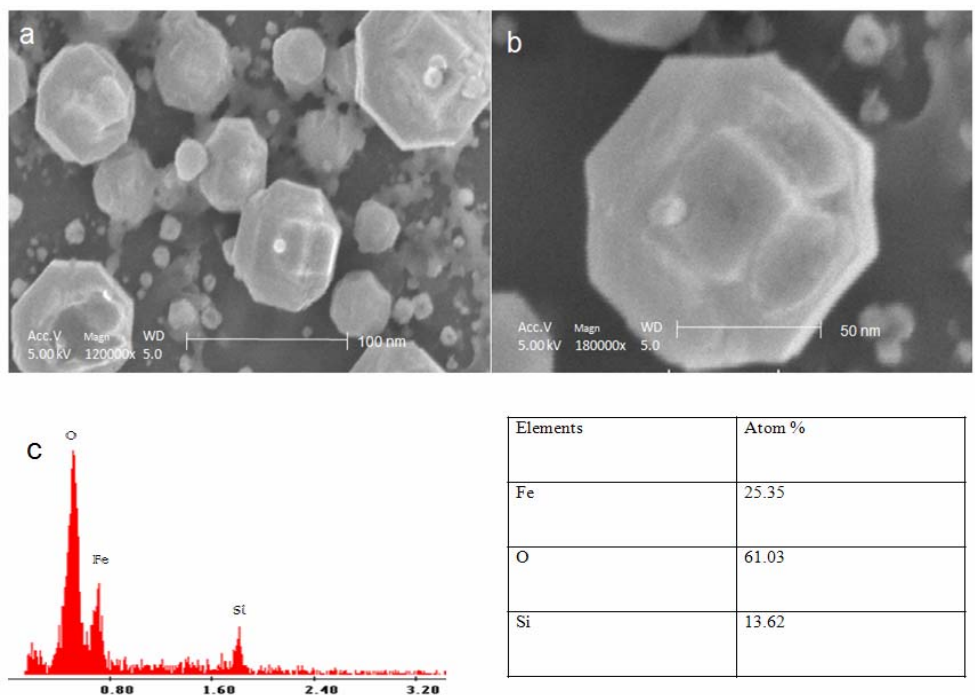

**Figure 2:** FE-SEM image of the Fe$_2$O$_3$ polyhedral nanocrystals obtained at 1000 °C annealing temperature. (a) The nanocrystals are polyhedral with different smooth surfaces. (b) High resolution FE-SEM image clearly indicating different faces of polyhedral nanocrystal. (c) Represents the EDX spectrum with atomic percentage in table.

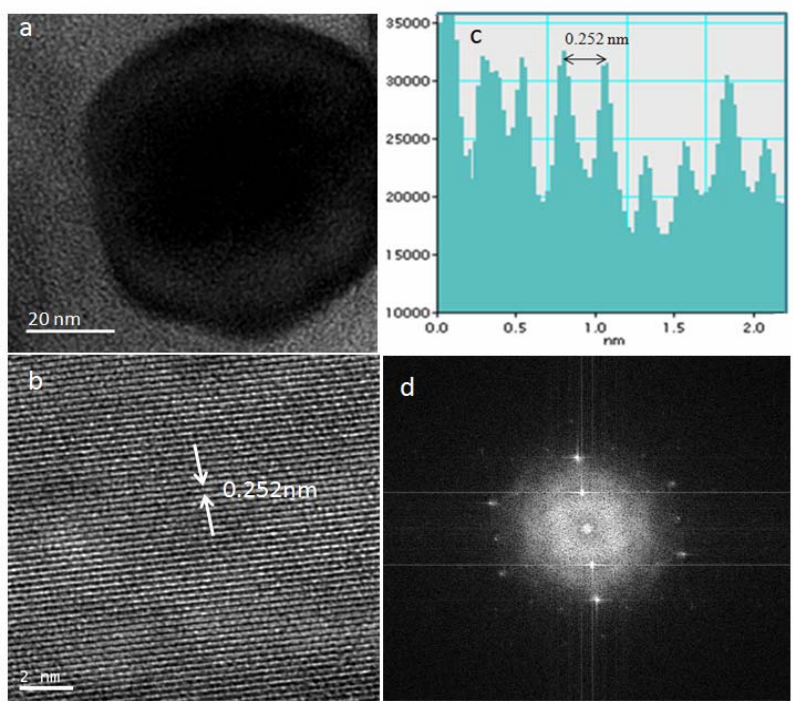

**Figure 3:** (a) TEM image, a Fe$_2$O$_3$ polyhedral nanocrystal. (b) High resolution TEM image showing parallel 110 planes. (c) Atomic profile for the HRTEM image. (d) Selected area diffraction pattern (SADP) of the nanocrystal.